\documentclass[11pt,a4paper]{article}
\usepackage{jheppub}
\usepackage[utf8]{inputenc}
\usepackage[english]{babel}

\usepackage{amsfonts}

\usepackage{tcolorbox}[most]

\newcommand{\eps}{\varepsilon}
\newcommand{\dx}{\mathrm{d}}
\newcommand{\ghy}{{}_2\mathrm{F}_1}
\newcommand{\Beta}{\mathrm{B}}

\newcommand{\approxLP}{\overset{\text{LP}}{\approx}}
\newcommand{\regop}[1]{\mathcal{M}_{#1}}
\newcommand{\clp}[1]{C_{#1}^\text{LP}}

\title{Expansion by regions meets angular integrals}

\author[a,b]{Vladimir A. Smirnov}
\author[c]{and Fabian Wunder}

\affiliation[a]{Skobeltsyn Institute of Nuclear Physics, Moscow State University, \\ Leninskie gory, 119992 Moscow, Russia}
\affiliation[b]{Moscow Center for Fundamental and Applied Mathematics, 119992 Moscow, Russia} 
\affiliation[c]{Institute for Theoretical Physics, University of T\"ubingen, \\ 
	Auf der Morgenstelle 14, 72076 T\"ubingen, Germany}

\emailAdd{smirnov@theory.sinp.msu.ru}
\emailAdd{fabian.wunder@uni-tuebingen.de}

\abstract{}
\arxivnumber{}

\abstract{
We study the small-mass asymptotic behavior of so-called angular integrals, appearing in phase-space calculations in perturbative quantum field theory.
For this purpose we utilize the strategy of expansion by regions, which is a universal method both for multiloop Feynman integrals and various parametric integrals.
To apply the technique to angular integrals, we convert them into suitable parametric integral representations,
which are accessible to existing automation tools.
We use the the code {\tt asy.m} to reveal regions contributing to the asymptotic expansion of angular integrals.
To evaluate the contributions of these regions in an epsilon expansion we apply the method of Mellin-Barnes representation.
Our approach is checked against existing results on angular integrals revealing a connection between contributing regions and angular integrals constructed from an algebraic decomposition.
We explicitly calculate the previously unknown asymptotics for angular integrals with three and four denominators and formulate a conjecture for the leading asymptotics and the pole part for a general number of denominators and masses.}

\begin{document}
\maketitle

\flushbottom

\section{Introduction}
\label{sec: Introduction}

Angular integrals \cite{Schellekens:1981,vanNeerven:1985,Beenakker:1988,Somogyi:2011,Lyubovitskij:2021,Wunder:2024} appear in a plethora of phase-space calculations in perturbative quantum field theory \cite{Bolzoni:2010,Anastasiou:2013, Lillard:2016,Kotlarski:2016,Lionetti:2018, Specchia:2018,Bahjat-Abbas:2018, Baranowski:2020,Blumlein:2020,Isidori:2020,Alioli:2022,Assi:2023,Catani:2023,Pal:2023,Devoto:2024,Rowe:2024}.
In the context of Quantum Chromodynamics   they found application in theoretical predictions for deep-inelastic scattering   \cite{Duke:1982,Hekhorn:2019}, semi-inclusive deep-inelastic scattering   \cite{Anderle:2016,Wang:2019}, the Drell-Yan process   \cite{Matsuura:1989,Matsuura:1990,Hamberg:1991,Mirkes:1992,Bahjat-Abbas:2018}, hadron-hadron scattering \cite{Ellis:1980}, heavy quark production \cite{Beenakker:1988}, prompt-photon production \cite{Gordon:1993,Rein:2024}, and single-spin asymmetries \cite{Schlegel:2012, Ringer:2015}.

To regularize collinear singularities, caused by the presence of massless particles, angular integrals need to be calculated in $d=4-2\eps$ dimensions \cite{tHooft:1972,Bollini:1972}.
In the cases of two and fewer denominators, the $\eps$-expansion of these integrals is well understood.
Here, results to all orders are available \citep{Lyubovitskij:2021}.
Furthermore, analytic results in terms of hypergeometric functions are known in these cases \cite{Beenakker:1988,Somogyi:2011,Lyubovitskij:2021}.
For three and more denominators the literature is much sparser.
Explicit $\eps$-expansions have only been published for the massless three-denominator integral  up to finite order in $\eps$ \cite{Somogyi:2011}.
Beyond that only general Mellin-Barnes representations are available \cite{Somogyi:2011}. 

The purpose of this work is two-fold.
First, we want to amend the knowledge about angular integrals in the multi-denominator case by studying the asymptotic behavior of angular integrals in the massless limit.
For up to two denominators results in this direction were presented recently in \cite{Wunder:2024}.
The second aim of this work is to present a novel application of the well-known method of expansion by regions~\cite{Beneke:1997zp,Smirnov:2002pj,Smirnov:2012gma}, developed in the context of (multi-)loop integrals, to a new class of integrals.
The expansion by regions method has been reformulated in the language of the Feynman parametric representation in \cite{Smirnov:1999bza}, for a recent review see~\cite{Smirnov:2021dkb}.
Expansion by regions can be applied not only to Feynman integrals and the corresponding parametric representations but also to much more general parametric integrals. 

To calculate the asymptotic expansion in the massless limit we adopt the following strategy.
We start by bringing the angular integrals into the form of parametric integrals suitable for expansion by regions.
From there we can proceed automatically, with the help of the corresponding computer code {\tt asy.m} \cite{Pak:2010pt,Jantzen:2012mw}, which is based on the geometry of polytopes and is also available with the
\texttt{FIESTA} distribution package \cite{FIESTA3:2014,FIESTA4:2016}.
We find one region associated with each of the masses, which gives a rather simple contribution, together with a single more complicated contribution associated with the massless limit.
This decomposition in regions is then compared with an algebraic decomposition used in \cite{Wunder:2024}.
After expanding the contribution of each region in a Laurent series in $\eps$, the poles in $\eps$ cancel as expected.
Where possible, consistency checks with known identities of angular integrals are performed.

Results of this work include parametric integral representations for $n$-denominator angular integrals applicable to the expansion by regions code \texttt{asy.m} together with the explicit calculation of the previously unknown asymptotics for angular integrals with three and four denominators.
Based on these findings we formulate a conjecture for the leading asymptotics and the pole structure for general $n$ denominators and $m$ masses.
 
This manuscript is organized as follows.
In the remainder of Section~\ref{sec: Introduction} we give the definition of angular integrals in $d$ dimensions, recall important reduction identities, state the kind of integral representation we are after for use in the program \texttt{asy.m}, and provide a brief summary of the expansion by regions prescription.
In Section~\ref{sec: Integral representations of angular integrals} we derive parametric integral representations for angular integrals of the form suitable for expansion by regions.
In Section~\ref{sec: Evaluating leading power asymptotics with asy} these integral representations are used to establish the asymptotic expansion of angular integrals in the massless limit with \texttt{asy.m}.
Section \ref{sec: Conclusion} concludes the paper.

\subsection{Definition of angular integrals}
\label{sec: Angular integrals}
The notation employed for angular integrals follows the references \cite{Somogyi:2011, Lyubovitskij:2021}.
Angular integrals in $d$ dimensions are defined as
\begin{align}
\Omega_{j_1,j_2,\dots,j_n}(v_1,\dots,v_n;d)\equiv\int\dx\Omega_{d-1}(k) \prod_{i=1}^n\frac{1}{(v_i\cdot k)^{j_i}}
\end{align}
with normalized $d$-vectors
\begin{align}
v_i=(1,\mathbf{v}_i)\,,\quad
k=(1,\mathbf{k})=(1,\dots,\cos\theta_n\prod_{i=1}^{n-1}\sin\theta_i,\dots,\cos\theta_2\sin\theta_1,\cos\theta_1)
\,.
\end{align}
For the $v_i$ one may choose an explicit parametrization of the form
\begin{align}
v_1&=(1,\mathbf{0}_{d-2},\beta_1)\,,\\
v_2&=(1,\mathbf{0}_{d-3},\beta_2\sin\chi_2^{(1)},\beta_2\cos\chi_2^{(1)})\,,\\
v_3&=(1,\mathbf{0}_{d-4},\beta_3\sin\chi_3^{(1)}\sin\chi_3^{(2)},\beta_3\sin\chi_3^{(1)}\cos\chi_3^{(2)},\beta_3\cos\chi_3^{(1)})\,,\\
&\vdots\nonumber\\
v_n&=(1,\mathbf{0}_{d-1-n},\beta_n \prod_{i=1}^{n-1}\sin\chi_n^{(i)},\beta_n \cos\chi_n^{(n)}\prod_{i=1}^{n-2}\sin\chi_n^{(i)},\dots,\beta_n\sin\chi_n^{(1)}\cos\chi_n^{(2)},\beta_n\cos\chi_n^{(1)})\,,
\end{align}
even though, in many cases, it is more convenient to keep the propagators in parametrization-free form, which preserves the explicit rotational invariance.
The angular integration measure is given by
\begin{align}
\dx\Omega_{d-1}(k) = \prod_{i=1}^{n}\dx\theta_i\,\sin^{2-i-2\eps}\theta_i\,\dx\Omega_{d-1-n}(k)\,,
\end{align}
the integration bound for each $\theta_i$ is $0\leq\theta_i\leq \pi$.
The $d-1-n$ dimensional part is transverse to all $v_i$ in the integrand, hence it can be integrated out as
\begin{align}
\Omega_{d-1-n}=\int\dx\Omega_{d-1-n}(k)=\frac{2 \pi^\frac{d-n}{2} }{\Gamma\left(\frac{d-n}{2}\right)}\,.
\end{align}
In the literature, e.g. \cite{Somogyi:2011,Lyubovitskij:2021,vanNeerven:1985,Beenakker:1988, Wunder:2024}, a commonly used normalization for angular integrals in $d=4-2\eps$ dimensions, especially those with only a few propagators, is
\begin{align}
I_{j_1,\dots,j_n}(v_1,\dots,v_n;\eps)\equiv\frac{\Omega_{j_1,j_2,\dots,j_n}(v_1,\dots,v_n;4-2\eps)}{\Omega_{-2\eps}}
\end{align}
with
\begin{align}
\Omega_{-2\eps}=2^{1-2\eps}\pi^{-\eps}\frac{\Gamma(1-2\eps)}{\Gamma(1-\eps)}.
\end{align}
This normalization simplifies the $\eps$ expansion by removing factors of the Euler-Mascheroni constant $\gamma_\text{E}$.

The angular integrals depend on the kinematic invariants $v_{ij}\equiv v_i\cdot v_j$.
If $v_{ii}=0$, we call the $i$th propagator \textit{massless}, else \textit{massive}.
A useful common notation is to indicate the number of non-zero masses in the integral with an explicit label, i.e. $I_{j_1,\dots,j_n}^{(m)}$ for an integral with $m$ masses.
The parametric variables $\beta_i$ and $\chi_j^{(i)}$, where $1\leq i\leq n$ and $i<j\leq n$, can be expressed through invariant variables $v_{ij}$ via
\begin{align}
\beta_i&=\sqrt{1-v_{ii}}\,,\;
\cos\chi_j^{(1)}=\frac{1-v_{1j}}{\beta_1\beta_j}\,,\;
\cos \chi_j^{(2)}=\frac{1-v_{2j}-\beta_2\beta_j\cos\chi_2^{(1)}\cos\chi_j^{(1)}}{\beta_2\beta_j\sin\chi_2^{(1)}\sin\chi_j^{(1)}}\,,
\nonumber\\
\cos \chi_j^{(3)}&=\frac{1-v_{3j}-\beta_3\beta_j\left(\cos\chi_3^{(1)}\cos\chi_j^{(1)}+\sin\chi_3^{(1)}\sin\chi_j^{(1)}\cos\chi_3^{(2)}\cos\chi_j^{(2)}\right)}{\beta_3\beta_j\sin\chi_3^{(1)}\sin\chi_j^{(1)}\sin\chi_3^{(2)}\sin\chi_j^{(2)}}\,,\dots
\end{align}
\subsection{Reduction identities for angular integrals}
\label{sec:Reduction identities for angular integrals}
We observe that for establishing the asymptotic behavior of any angular integral with integer propagator powers $j_1,\dots ,j_n$ it is sufficient to determine the behavior of the master integrals after IBP reduction, see \cite{Lyubovitskij:2021} for the two denominator case.
The master integrals are those with $j_i=0,1$ for $i=1,\dots,n$, i.e. there is only the master integral $I_{1,\dots,1}$ with manifestly $n$ denominators.
Therefore it suffices to derive the asymptotic behavior for $I_{1,\dots,1}$ in section \ref{sec: Evaluating leading power asymptotics with asy}. 

Furthermore, any master integral with linearly dependent vectors $v_1,\dots,v_n$ can be reduced to $n-1$-denominator angular integrals by partial fractioning identities.
Assuming the fixed vectors $v_i$ are all in a four dimensional subspace, we have a maximum of three linearly dependent denominators, hence, under this assumption, all higher denominator angular integrals can be reduced to the three denominator case.
We will not assume linear dependence in the following, however we will use this identity as a cross-check between the three- and 
four-denominator angular integrals in section \ref{sec: Four denominators asy}.

Another important identity for angular integrals is the \textit{two-point splitting lemma} \cite{Lyubovitskij:2021,Wunder:2024}, which allows to rewrite the product of two massive denominators as a sum of two products of one massive and one auxiliary massless denominator.
An iterative use allows for writing an angular integral with any number of masses as a sum of integrals with only one non-zero mass.
Therefore, the general integral $I_{1,\dots,1}$ can be expressed through a sum of $I_{1,\dots,1}^{(0)}$ and $I_{1,\dots,1}^{(1)}$.
This was heavily used in \cite{Wunder:2024}, however for expansion by regions it is less useful since, as we will see explicitly in section \ref{sec: Two denominators asy}, the splitting may change the scaling behavior of the angular integral.
Additionally, keeping all $n$ masses has the benefit that the cancellation of poles between regions in the fully massive case serves as a useful consistency check.

The asymptotics in the massless limit is known for up to two denominators \cite{Wunder:2024}.
For three and more denominators, much less is known in this regard.
Explicit $\eps$-expansions have been published only for the massless three-denominator integral to finite order in $\eps$ \cite{Somogyi:2011}.
In this work we will add the leading small mass asymptotics to all orders in $\eps$ to the three-denominator result and derive the asymptotic behavior for four-denominators.
Based on these findings we formulate a conjecture for the pole part for the general case of $n$ denominators and $m$ masses.

\subsection{Integral representations compatible with \texttt{asy.m}}
\label{sec: Integral representations compatible with asy}

The central idea of the present work is an application of the strategy of expansion by regions to angular integrals to obtain the asymptotic behavior of angular integrals in $d=4-2\eps$ dimensions in the massless limit.
This method and  the corresponding computer code \texttt{asy.m} has been developed in the context of multiloop Feynman integrals, but is also applicable to certain classes of parametric integrals, for a recent example see \cite{Belitsky:2023}, 
where the method was applied to an off-shell Wilson loop.
Of course, the code can work with parametric representations which have the same form as the Feynman parametric representation of Feynman  integrals, i.e.
where the integrand is a product of two polynomials raised to powers which are linear in $d$ with a delta function 
$\delta\left(\sum x_i -1\right)$ where the summation can be chosen in a convenient way according to the folklore Cheng-Wu theorem.\footnote{For a proof see e.g. \cite{Smirnov:2012gma}, p. 46.}
Note that there is no direct analog of the Cheng-Wu theorem for angular integrals since the normalization of the zero-component of the $v_i$ vectors to $1$ breaks the homogeneity of the corresponding Feynman representation.
This is however not a problem since \texttt{asy.m} also works on various other integral representations.

In this paper, we apply it to $n$-fold parametric integrals where each of the integration variables $t_1,\dots t_n$ runs from zero to infinity.
The integrand is supposed to be a product of polynomials $P_i$ raised to some powers $a_i+\eps\,b_i$.
The $P_i$ are polynomials in the integration variables with coefficients depending on a set of kinematic variables $s_1,\dots,s_m$.
The exponent coefficients $a_i$ and $b_i$ are numbers, in most practical cases integers or half-integers.

This means that we are looking for an integral representation of the form
\begin{align}
I=\int_0^\infty\dx t_1\,\dots\int_0^\infty\dx t_n\,\prod_{i=1}^N P_i^{a_i+\eps b_i}(t_1,\dots,t_n)
\label{eq: parametric integral representation}
\end{align}
for angular integrals.

\subsection{Expansion by regions in a nutshell}
\label{sec: Expansion by regions in a nutshell}

As already mentioned above, expansion by regions has been first developed for Feynman integrals.
The original formulation  \cite{Beneke:1997zp} was based on analyzing various regions in loop momentum space, performing the corresponding expansions in small parameters within each of them, dropping out restrictions on the integration domain specifying the individual regions, and finally summing over the contributions of all regions under consideration.
Practically, one didn't bother about possible double (and multiple) counting when collecting contributions of regions.\footnote{with one exception: see \cite{Jantzen:2011} where
regions were introduced in an `honest' way, with specifying integration domains by explicit inequalities. Here interplay of 
various regions (in its primary sense) was demonstrated explicitly using some characteristic two-loop examples.}
Really, the notion of {\em region} was understood in many practical calculations as a scaling of loop integration momenta measured in powers of an expansion parameter.
However, to reveal all the relevant regions/scalings, was, in most cases, a highly non-trivial task.
It was the formulation of expansion by regions \cite{Smirnov:1999bza,Pak:2010pt} in terms of parametric integrals which opened a way to systematically solve this task. 
The geometric formulation gave also the possibility of a generalization of the method to a wider class of integrals such as the parametric integrals discussed in section \ref{sec: Integral representations compatible with asy}.

A recent review on the method of expansion by regions can be found in \cite{Smirnov:2021dkb}, for some recent applications see \cite{Ananthanarayan:2019,Ananthanarayan:2020,Heinrich:2022,terHoeve:2023ehm,Niggetiedt:2023uyk,Beneke:2023wmt,Ma:2023hrt,Gardi:2022khw}.
Expansion by regions is now implemented not only with {\tt FIESTA} but also within the package {\tt pySecDec} -- see~\cite{Heinrich:2022}.
Here we just want to give a very brief overview of the prescriptions as applied to parametric integral, to introduce some notation used in this work.

The prescription of expansion by regions \cite{Pak:2010pt,Jantzen:2012mw,Semenova:2019,Smirnov:2021dkb} states that and $n$-fold parametric integral $I$ of the structure of Eq.~\eqref{eq: parametric integral representation} with polynomials depending on a small parameter $y$ has an asymptotic expansion in $y$ of the form
\begin{align}
I\sim\sum_{r_i\in r}\regop{r_i} I,
\end{align}
where the sum runs over a set $r$ of contributing regions $r_i$.
These regions $r_i$, which correspond to facets of Newton polytopes associated with the integrand polynomials, are vectors with $n$ entries each.
For the purpose of this work the set $r$ will be determined automatically by \texttt{asy.m}.
The operator $\regop{r_i}$ associated with each region acts on the integrand of $I$ by scaling $t_j \to y^{(r_i)_j} t_j$, multiplying by  $y^{\sum_{j=1}^{n} (r_i)_j}$, expanding in $y$ at $y=0$ and setting $y=1$ in the end.
If necessary additional regulators $t_j^{\eta_j}$ are introduced to ensure convergence of the integrals.
If there are no poles in the regulators after analytic continuation to $\eta_j=0$, they can be dropped directly.
This will be the case for all integrals under consideration in this work.
Generally, if poles in the regulators appear, these will cancel after combining regions.

The small parameter $y$ in this work is associated with small masses $v_{ii}$ in the angular integrals.
It will be introduced by multiplying each $v_{ii}$ with an explicit factor of $y$.
For simplicity, we introduce the notation $\clp{n,i}$ to refer to the leading power contribution from region $r_i$ to the $n$-denominator angular integral $I_{1,\dots,1}$, i.e.
\begin{align}
\regop{r_i} I_{\underbrace{1,\dots,1}_n}\approxLP \clp{n,i},
\end{align}
where ``$\approxLP$" denotes the leading power approximation in the masses $v_{ii}$.
With this notation the leading asymptotics of the integral $I_{1,\dots,1}$ is given by
\begin{align}
I_{1,\dots,1}\approxLP \sum_i \clp{n,i}\,,
\end{align}
where the sum runs over all regions detected by \texttt{asy.m}.
\section{Parametric integral representations for angular integrals}
\label{sec: Integral representations of angular integrals}
For the construction of suitable integral representations, we start with the one denominator case.
From there we successively increase the number of denominators.
Using Feynman parametrization we can systematically construct representations for more denominators, with
the integral representation of the one-denominator integral serving as the main building block.
\subsection{Integral representation for one denominator}
Writing the one-denominator integral
\begin{align}
I_{j_1}(v_{11};\eps)=\int\frac{\dx\Omega_{3-2\eps}(k)}{\Omega_{-2\eps}}\,\frac{1}{(v_1\cdot k)^{j_1}}
\end{align}
in explicit spherical coordinates with $v_1$ pointing in the $z$ direction, it is \cite[Eq.~(3.26)]{Lyubovitskij:2021}
\begin{align}
I_{j_1}(v_{11};\eps)=\int_0^\pi\dx\theta_1\frac{\sin^{1-2\eps}\theta_1}{(1-\beta\cos\theta_1)^j}\int_0^\pi\dx\theta_2\sin^{-2\eps}\theta_2\,,
\end{align}
with $\beta=\sqrt{1-v_{11}}$.
Substituting $\cos\theta_i=1-2 t_i$ one obtains an Euler type integral representation \cite[Eq.~(3.28)]{Lyubovitskij:2021}
\begin{align}
I_{j_1}(v_{11};\eps)=\frac{2^{1-4\eps}}{(1-\beta)^{j_1}}\frac{\Gamma^2\!\left(\frac{1}{2}-\eps\right)}{\Gamma(1-2\eps)}\int_0^1\dx t_1\,t_1^{-\eps}(1-t_1)^{-\eps}\left(1+\frac{2\beta}{1-\beta}\,t_1\right)^{-j_1}\,,
\end{align}
which can be expressed in terms of a Gauss hypergeometric function as \cite[Eq.~(3.30)]{Lyubovitskij:2021}
\begin{align}
I_{j_1}(v_{11};\eps)=\frac{I_0(\eps)}{(1-\beta)^{j_1}}\,\ghy\!\left(j_1,1-\eps,2-2\eps;-\frac{2\beta}{1-\beta}\right),
\label{eq: One-denominator angular integral hypergeometric 1}
\end{align}
where $I_0(\eps)=\frac{2\pi}{1-2\eps}$.

The Gauss hypergeometric function $\ghy$ admits for the integral representation (see \cite[Eq.~(15.6.2{\_}5)]{NIST:DLMF})
\begin{align}
\ghy(a,b,c;z)=\frac{\Gamma(c)}{\Gamma(b)\Gamma(c-b)}\int_0^\infty\dx t\,t^{b-1}\,(1+t)^{a-c}\,(1+(1-z)\,t)^{-a}\,.
\label{eq: 2F1 integral representation for asym}
\end{align}
Hence, we have
\begin{align}
I_{j_1}(v_{11};\eps)=2\pi\,\frac{\Gamma(1-2\eps)}{\Gamma^2(1-\eps)}\int_0^\infty\dx t\,t^{-\eps}\,(1+t)^{j_1-2+2\eps}\,(1-\beta+(1+\beta)\,t)^{-{j_1}}\,,
\label{eq: One-denominator angular integral asym prelim}
\end{align}
which is of the desired form for use with \texttt{asy.m}.

However, we observe that the expression contains a square root $\beta=\sqrt{1-v_{11}}$ of the mass parameter.
This complicates matters when we try to use the integral representation of the one denominator integral within a bigger integral, since the ``mass" will depend on additional integration variables.
Fortunately, we can get rid of these square roots for the one-denominator integral by using the quadratic transformation (see \cite[Eq.~
(15.8.13)]{NIST:DLMF})
\begin{align}
\ghy(a,b,2b;z)=\left(1-\frac{z}{2}\right)^{-a}\ghy\!\left(\frac{a}{2},\frac{a+1}{2},b+\frac{1}{2}\left(\frac{z}{2-z}\right)^2\right)
\end{align}
on the hypergeometric function in Eq.~\eqref{eq: One-denominator angular integral hypergeometric 1}.
This gives \cite[Eq.~(3.33)]{Lyubovitskij:2021}
\begin{align}
I_{j_1}(v_{11};\eps)=I_0(\eps)\,\ghy\!\left(\frac{j_1}{2},\frac{j_1+1}{2},\frac{3}{2}-\eps;1-v_{11}\right)\,.
\end{align}
Using the integral representation \eqref{eq: 2F1 integral representation for asym} on this form of the hypergeometric function, results in
\begin{align}
I_{j_1}(v_{11};\eps)=\frac{2\pi}{1-2\eps}\,\frac{\Gamma\!\left(\frac{3}{2}-\eps\right)}{\Gamma\!\left(\frac{j_1+1}{2}\right)\Gamma\!\left(1-\frac{j_1}{2}-\eps\right)}\int_0^\infty\dx t\, t^\frac{j_1-1}{2}\,(1+t)^{\frac{j_1-3}{2}+\eps}\,(1+v_{11} t)^{-\frac{j_1}{2}}\,,
\label{eq: One-denominator angular integral asym}
\end{align}
which again is of a form compatible with \texttt{asy.m}.
Effectively, compared with the alternative representation from Eq.~\eqref{eq: One-denominator angular integral asym prelim}, we have interchanged the square roots in the kinematic variables for half-integers in the exponents.

By using the symmetry of $\ghy$ in its first two arguments and interchanging the roles of $a$ and $b$ in Eq.~\eqref{eq: 2F1 integral representation for asym} we also have the further equivalent representation
\begin{align}
I_{j_1}(v_{11};\eps)=\frac{2\pi}{1-2\eps}\,\frac{\Gamma\!\left(\frac{3}{2}-\eps\right)}{\Gamma\!\left(\frac{j_1}{2}\right)\Gamma\!\left(\frac{3-j_1}{2}-\eps\right)}\int_0^\infty\dx t\, t^{\frac{j_1}{2}-1}\,(1+t)^{\frac{j_1}{2}-1+\eps}\,(1+v_{11} t)^{-\frac{j_1+1}{2}}\,.
\label{eq: One-denominator angular integral asym alternative}
\end{align}
Depending on whether $j_1$ is even or odd the square roots are distributed differently in the two forms Eq.~\eqref{eq: One-denominator angular integral asym} and \eqref{eq: One-denominator angular integral asym alternative}.

The integral representations \eqref{eq: One-denominator angular integral asym} and \eqref{eq: One-denominator angular integral asym alternative} will be the cornerstones for building up suitable integral representations for angular integrals with more denominators.
Note that we used the known full hypergeometric result for the one-denominator integral in deriving these integral representation.
Hence, in this case, the application of expansion by regions to extract the asymptotic expansion might seem circular.
Let us emphasize that for the following cases of multiple denominators no additional analytic input is required and we are able to establish integral representations also for the cases where a hypergeometric representation is not known.

\subsection{Integral representation for two denominators}
Now we consider the angular integral with two denominators,
\begin{align}
I_{j_1,j_2}(v_{12},v_{11},v_{22};\eps)=\int\frac{\dx\Omega_{3-2\eps}(k)}{\Omega_{-2\eps}}\,\frac{1}{(v_1\cdot k)^{j_1}(v_2\cdot k)^{j_2}}\,.
\end{align}
The key idea for extending the result for the one-denominator integral from the previous section to more denominators is a \textit{Feynman parametrization} of the propagators.
This gives \cite[Eq.~(3.69)]{Lyubovitskij:2021}
\begin{align}
I_{j_1,j_2}(v_{12},v_{11},v_{22};\eps)&=\frac{1}{\Beta(j_1,j_2)}\int_0^1\dx x_1\,x_1^{j_1-1}\int_0^1\dx x_2\,x_2^{j_2-1}\,\delta(1-x_1-x_2)\,I_{j_1+j_2}\!\left(v^2;\eps\right)\,,
\end{align}
where $v\equiv x_1 v_1+x_2 v_2$ and $\Beta(j_1,j_2)=\frac{\Gamma(j_1)\Gamma(j_2)}{\Gamma(j_1+j_2)}$ denotes the Euler Beta function.
The mass $v^2$ within the one-denominator angular integral is
\begin{align}
v^2=(x_1 v_1+x_2 v_2)^2=x_1^2 v_{11}+2 x_1 x_2 v_{12}+x_2^2 v_{22}^2\,.
\end{align}
Using the delta function to integrate out $x_2$, we obtain
\begin{align}
I_{j_1,j_2}(v_{12},v_{11},v_{22};\eps)&=\frac{1}{\Beta(j_1,j_2)}\int_0^1\dx x_1\,x_1^{j_1-1}(1-x_1)^{j_2-1}\,I_{j_1+j_2}\!\left(v^2;\eps\right)
\end{align}
where now $v^2=x_1^2 (v_{11}-2 v_{12}+v_{22})+2 x_1(v_{12}-v_{22})+v_{22}$.

The last step is to perform a substitution such that the integration domain extends from zero to infinity.
A suitable transformation is given by
\begin{align}
t=\frac{x_1}{1-x_1}\,,\qquad
x_1=\frac{t_1}{1+t_1}\,,\qquad
\dx x_1=\frac{\dx t_1}{(1+t_1)^2}\,,
\end{align}
resulting in
\begin{align}
I_{j_1,j_2}(v_{12},v_{11},v_{22};\eps)=\frac{1}{\Beta(j_1,j_2)}\int_0^\infty&\dx t_1\, t_1^{j_1-1}(1+t_1)^{-j_1-j_2}\nonumber\\
\times&I_{j_1+j_2}\!\left(\frac{t_1^2 v_{11}+2 t_1 v_{12}+v_{22}}{(1+t_1)^2};\eps\right).
\end{align}
Now we can plug in the integral representation \eqref{eq: One-denominator angular integral asym} to obtain
\begin{align}
&I_{j_1,j_2}(v_{12},v_{11},v_{22};\eps)=\frac{2\pi}{1-2\eps}\,\frac{\Gamma\!\left(\frac{3}{2}-\eps\right)}{\Beta(j_1,j_2)\Gamma\!\left(\frac{j_1+j_2+1}{2}\right)\Gamma\!\left(1-\frac{j_1+j_2}{2}-\eps\right)}\nonumber\\
&\quad\times\int_0^\infty\dx t_1\int_0^\infty\dx t_2\, t_1^{j_1-1}\, t_2^\frac{j_1+j_2-1}{2}\,(1+t_2)^{\frac{j_1+j_2-3}{2}+\eps}\nonumber\\
&\qquad\qquad\qquad\times\left[(1+t_1)^2+(t_1^2 v_{11}+2 t_1 v_{12}+v_{22}) t_2\right]^{-\frac{j_1+j_2}{2}}\,.
\label{eq: Two-denominator angular integral asym}
\end{align}
This is of the form required by \texttt{asy.m}.

Using \eqref{eq: One-denominator angular integral asym alternative} instead of \eqref{eq: One-denominator angular integral asym} for the one-denominator integral gives an alternative representation:
\begin{align}
&I_{j_1,j_2}(v_{12},v_{11},v_{22};\eps)=\frac{2\pi}{1-2\eps}\,\frac{\Gamma\!\left(\frac{3}{2}-\eps\right)}{\Beta(j_1,j_2)\Gamma\!\left(\frac{j_1+j_2}{2}\right)\Gamma\!\left(\frac{3-j_1-j_2}{2}-\eps\right)}\nonumber\\
&\quad\times\int_0^\infty\dx t_1\int_0^\infty\dx t_2\, t_1^{j_1-1}\,(1+t_1)\, t_2^{\frac{j_1+j_2}{2}-1}\,(1+t_2)^{\frac{j_1+j_2}{2}-1+\eps}\nonumber\\
&\qquad\qquad\qquad\times\left[(1+t_1)^2+(t_1^2 v_{11}+2 t_1 v_{12}+v_{22}) t_2\right]^{-\frac{j_1+j_2+1}{2}}\,.
\label{eq: Two-denominator angular integral asym alternative}
\end{align}
\subsection{Integral representations for three and more denominators}
The method employed for the two-denominator angular integral straightforwardly generalizes to three and more denominators.
Here, we will explicitly derive the integral representation for three denominators and afterwards give the necessary steps for the general case of $n$ denominators.

The three-denominator angular integral
\begin{align}
I_{j_1,j_2,j_3}(v_{12},v_{13},v_{23},v_{11},v_{22},v_{33};\eps)=\int\frac{\dx \Omega_{3-2\eps}(k)}{\Omega_{-2\eps}}\,\frac{1}{(v_1\cdot k)^{j_1}(v_2\cdot k)^{j_2}(v_3\cdot k)^{j_3}}
\end{align}
admits for the Feynman parametrization
\begin{align}
&I_{j_1,j_2,j_3}(v_{12},v_{13},v_{23},v_{11},v_{22},v_{33};\eps)=\frac{1}{\Beta(j_1,j_2,j_3)}
\int_0^1\dx x_1\int_0^1\dx x_2\int_0^1\dx x_3\,x_1^{j_1-1}x_2^{j_2-1}x_3^{j_3-1}
\nonumber\\
&\qquad\qquad\qquad\qquad\qquad\times\delta(1-x_1-x_2-x_3)\,I_{j_1+j_2+j_3}\!\left((x_1 v_1+x_2 v_2+x_3 v_3)^2;\eps\right).
\end{align}
To map the integration domain onto integrals from zero to infinity, we can use the delta function to evaluate the $x_3$ integral and subsequently use the transformation
\begin{align}
&t_1=\frac{x_1}{1-x_1}\,,\quad
t_2=\frac{x_2}{1-x_1-x_2}\,,\quad
x_1=\frac{t_1}{1+t_1}\,,\quad
x_2=\frac{t_2}{(1+t_1)(1+t_2)}\,,\nonumber\\
\end{align}
with Jacobian $ (1+t_1)^{-3}(1+t_2)^{-2}$,
on the remaining two-fold integral.
Under this transformation, the Feynman integration measure becomes
\begin{align}
\int_0^1\dx x_1\int_0^1\dx x_2\int_0^1\dx x_3\,x_1^{j_1-1}\,x_2^{j_2-1}\,x_3^{j_3-1}\delta(1-x_1-x_2-x_3)
\nonumber\\
=
\int_0^\infty\dx t_1\int_0^\infty \dx t_2\,t_1^{j_1-1}(1+t_1)^{-j_1-j_2-j_3}t_2^{j_2-1}(1+t_2)^{-j_2-j_3}\,.
\end{align}
The mass of the remaining one denominator integral expressed in the new variables $t_1$, $t_2$ is
\begin{align}
(x_1 v_1+x_2 v_2+x_3 v_3)^2=\frac{t_1^2(1+t_2)^2 v_{11}+2 t_1 (1+t_2)(v_{13}+t_2 v_{12})+t_2^2 v_{22}+2 t_2 v_{23}+v_{33}}{(1+t_1)^2(1+t_2)^2}\,.
\end{align}
Plugging in the integral representation of the one-denominator angular integral \eqref{eq: One-denominator angular integral asym}, we obtain the following representation for the three denominator angular integral:
\begin{align}
&I_{j_1,j_2,j_3}(v_{12},v_{13},v_{23},v_{11},v_{22},v_{33};\eps)=\frac{2\pi}{1-2\eps}\frac{\Gamma\!\left(\frac{3}{2}-\eps\right)}{\Beta(j_1,j_2,j_3)\Gamma\!\left(\frac{j+1}{2}\right)\Gamma\!\left(1-\frac{j}{2}-\eps\right)}
\nonumber\\
&\times
\int_0^\infty\dx t_1\int_0^\infty \dx t_2\int_0^\infty \dx t_3\,t_1^{j_1-1}t_2^{j_2-1}(1+t_2)^{j_1}t_3^\frac{j-1}{2}(1+t_3)^{\frac{j-3}{2}+\eps}\,w_3^{-\frac{j}{2}}(t_1,t_2,t_3)\,,
\label{eq: Three-denominator angular integral asym}
\end{align}
where $j=j_1+j_2+j_3$, $\Beta(j_1,j_2,j_3)=\frac{\Gamma(j_1)\Gamma(j_2)\Gamma(j_3)}{\Gamma(j)}$, and $w_3(t_1,t_2,t_3)$ denotes the polynomial
\begin{align}
w_3(t_1,t_2,t_3)&\equiv(1+t_1)^2(1+t_2)^2\nonumber\\
&+\left[t_1^2(1+t_2)^2 v_{11}+2 t_1 (1+t_2)(v_{13}+t_2 v_{12})+t_2^2 v_{22}+2 t_2 v_{23}+v_{33}\right]t_3.
\end{align}
The integral representation Eq.~\eqref{eq: Three-denominator angular integral asym} is of the form required by \texttt{asy.m}.

Using \eqref{eq: One-denominator angular integral asym alternative} instead of \eqref{eq: One-denominator angular integral asym} for the one-denominator integral gives an alternative representation:
\begin{align}
&I_{j_1,j_2,j_3}(v_{12},v_{13},v_{23},v_{11},v_{22},v_{33};\eps)=\frac{2\pi}{1-2\eps}\frac{\Gamma\!\left(\frac{3}{2}-\eps\right)}{\Beta(j_1,j_2,j_3)\Gamma\!\left(\frac{j}{2}\right)\Gamma\!\left(\frac{3-j}{2}-\eps\right)}
\nonumber\\
&\times
\int_0^\infty\dx t_1\int_0^\infty \dx t_2\int_0^\infty \dx t_3\,t_1^{j_1-1}(1+t_1)t_2^{j_2-1}(1+t_2)^{j_1+1}t_3^{\frac{j}{2}-1}(1+t_3)^{\frac{j}{2}-1+\eps}
\nonumber\\
&\qquad\qquad\qquad\qquad\qquad\times w_3^{-\frac{j+1}{2}}(t_1,t_2,t_3)\,,
\label{eq: Three-denominator angular integral asym alternative}
\end{align}
with the same abbreviations as in Eq.~\eqref{eq: Three-denominator angular integral asym}.
\\

For more than three denominators we can establish suitable integral representations analogously.
The necessary steps for the $n$-denominator angular integral are
\begin{itemize}
\item[1.] Use a Feynman parametrization with parameters $x_i$ for the denominator to write the $n$-denominator integral as an $n$-fold Feynman parameter integral over a one-denominator angular integral with the mass $v=\left(\sum_{i=1}^n x_i\right)^2$.
\item[2.] Use the delta function to integrate out $x_n$.
\item[3.] In the remaining $n-1$-fold integral, change variables to $t_i=\frac{x_i}{1-\sum_{l=1}^i x_l}$.
\item[4.] Write the mass $v$ in the new variables $t_1,\dots t_{n-1}$.
\item[5.] Plug in the integral representation \eqref{eq: One-denominator angular integral asym} or \eqref{eq: One-denominator angular integral asym alternative} for the one-denominator angular integral with mass $v$.
\item[6.] Collect and simplify the polynomial factors in the integrand.
\end{itemize} 
These steps lead to a $n$-fold integral representation of the $n$-denominator angular integral of the form required by \texttt{asy.m}.
Following the outlined algorithm and using Eq.~\eqref{eq: One-denominator angular integral asym} in step 5 we find for the four denominator integral
\begin{align}
&I_{1,1,1,1}(v_{12},v_{13},v_{14},v_{23},v_{24},v_{34},v_{11},v_{22},v_{33},v_{44};\eps)=\frac{2\pi}{1-2\eps}\frac{\Beta(j_1,j_2,j_3,j_4)^{-1}\Gamma \left(\frac{3}{2}-\eps\right)}{\Gamma \left(\frac{j+1}{2}\right) \Gamma
   \left(-\eps-\frac{j}{2}+1\right)}
   \nonumber\\
   &
  \int_0^\infty\!\dx t_1\int_0^\infty\!\dx t_2\int_0^\infty\!\dx t_3\int_0^\infty\!\dx t_4\,t_1^{j_1-1} t_2^{j_2-1} (1+t_2)^{j_1} t_3^{j_3-1}
   (1+t_3)^{j_1+j_2} t_4^{\frac{j-1}{2}}
   (1+t_4)^{\eps+\frac{j-3}{2}} w_4^{-\frac{j}{2}}
   \label{eq: Four-denominator angular integral asym}
\end{align}
with
\begin{align}
w_4=(t_1+1)^2 (t_2+1)^2
   (t_3+1)^2+&t_4 \left[t_3^2 v_{33}+2 t_3 v_{34}+t_1^2 (t_2+1)^2
   (t_3+1)^2 v_{11}
   \right.\nonumber\\
   &\left.
   +2 t_1 (t_2+1) (t_3+1) (t_2 (t_3+1)
   v_{12}+t_3 v_{13}+v_{14})
   \right.\nonumber\\
   &\left.
   +t_2^2 (t_3+1)^2 v_{22}+2 t_2
   (t_3+1) (t_3 v_{23}+v_{24})+v_{44}\right]\;,
\label{w4}
\end{align}
where $j=j_1+j_2+j_3+j_4$ and $\Beta(j_1,j_2,j_3,j_4)=\frac{\Gamma(j_1)\Gamma(j_2)\Gamma(j_3)\Gamma(j_4)}{\Gamma(j)}$.

\subsection{Summary for parametric representations}
We explicitly constructed suitable integral representations for
\begin{itemize}
\item one-denominator angular integrals -- see eqs.~\eqref{eq: One-denominator angular integral asym}\,\&\,\eqref{eq: One-denominator angular integral asym alternative}\,,
\item two-denominator angular integrals -- see eqs.~\eqref{eq: Two-denominator angular integral asym}\,\&\,\eqref{eq: Two-denominator angular integral asym alternative}\,,
\item three-denominator angular integrals -- see eqs.~\eqref{eq: Three-denominator angular integral asym}\,\&\,\eqref{eq: Three-denominator angular integral asym alternative}\,,
\item four-denominator angular integrals -- see Eq.~\eqref{eq: Four-denominator angular integral asym}\,,
\end{itemize} 
and presented an algorithm to construct a corresponding representation for the $n$ denominator case.
In each case we can derive two suitable integral representations which differ in the distribution of integer and half-integer powers across the polynomial factors. 
In the following section these integral representations are used in \texttt{asy.m} to extract the asymptotic behavior of the angular master integrals, i.e. those with $j_1,\dots,j_n=1$, in the massless limit.

\section{Evaluating leading power asymptotics with asy.m}
\label{sec: Evaluating leading power asymptotics with asy}
We use the integral representations found in section \ref{sec: Integral representations of angular integrals} to evaluate the leading power asymptotics with the Mathematica code \texttt{asy.m}.
We start with the two-denominator master integral $I_{1,1}$ where the asymptotics is known from the literature.
Here, we find perfect agreement with the results from \cite{Wunder:2024}.
After this consistency check, we continue with the cases of three and four denominators, where we establish novel results.
\subsection{Asymptotics for two denominators}
\label{sec: Two denominators asy} 
Let us obtain the asymptotic behavior of the two-denominator integral $I_{1,1}$ using the integral representation from Eq.~\eqref{eq: Two-denominator angular integral asym} with the two indices $j_{1,2}$ equal to one,
\begin{align}
I_{1,1}=
\frac{4 \sqrt{\pi}\,\Gamma(3/2-\eps)}{(1-2\eps)\,\Gamma(-\eps)} 
\int_0^\infty\dx t_1\int_0^\infty \dx t_2 \,
\frac{(1+t_2)^{-1/2 + \eps}\sqrt{t_2}}{(1 + t_1)^2 + t_2 (t_1^2 v_{11} + 2 t_1 v_{12} + v_{22})}\;.
\label{I2}
\end{align}
In the massless limit it is $v_{11}\sim v_{22} \ll v_{12}$.
To reveal regions, i.e. scalings of integration variables which generate contributions to the expansion,
we apply the following command of \texttt{asy.m}:
\begin{verbatim}
r = WilsonExpand[(1 + t[2]) t[2], (1 + t[1])^2 
    + (v22 y + 2 v12 t[1] + v11 y t[1]^2) t[2], {t[1], t[2]}, {y -> x}]
\end{verbatim}
Here we introduced the small parameter of the problem, $y$, by multiplying it to all the small kinematic invariants.
In the first two arguments we have to put all the $t_i$-dependent polynomials which are raised to a power which is not a positive integer number.\footnote{See \cite{Jantzen:2012mw} for more details on the usage of \texttt{asy.m}.}
As an output of this command, we obtain 
\begin{align}
r =\{\{-1, -1\}, \{1, -1\}, \{0, 0\}\}.
\end{align}

According to the prescriptions of expansion by regions \cite{Pak:2010pt,Jantzen:2012mw,Smirnov:2021dkb} the asymptotics of $I_{1,1}$ is given by
\begin{align}
I_{1,1}\sim\sum_{r_i\in r}\regop{r_i} I_{1,1}.
\end{align}

Following the steps recalled in section \ref{sec: Expansion by regions in a nutshell}, we scale by $t_j \to y^{(r_i)_j} t_j$, multiply by  $y^{\sum_{j=1}^3 (r_j)_i}$, expand in $y$ at $y=0$ and set $y=1$ in the end.
We obtain for the leading power behavior of the first contribution
\begin{align}
\regop{r_1} I_{1,1}\approxLP\frac{4 \sqrt{\pi}\,\Gamma(3/2-\eps)}{(1-2\eps)\Gamma(-\eps)} 
\int_0^\infty\dx t_1\int_0^\infty \dx t_2 \,
\frac{t_1^{\eta_1-1}t_2^{\eps+\eta_2}}{t_1 + 2 v_{12} t_2 +  v_{11} t_1 t_2}\,,
\label{I2-1}
\end{align}
where $\eta_{1,2}$ are additional regulators to ensure convergence.
The integral in Eq.~\eqref{I2-1} can easily be evaluated by an application of a one-dimensional integration formula, with the following result
valid at general $\eps$ where the additional regulators $\eta_{1,2}$ have been put to zero:
\begin{align}
\clp{2,1}=\sqrt{\pi} \frac{\Gamma(1/2 - \eps) \Gamma(\eps)}{v_{12} v_{11}^\eps}=
\frac{\pi}{\eps}\left(\frac{v_{11}}{4}\right)^{-\eps}\frac{\Gamma(1+\eps)\Gamma(1-2\eps)}{v_{12}\,\Gamma(1-\eps)}\;.
\label{I2-1-res}
\end{align}
Since the regulators $\eta_i$ will always cancel in each region individually for the integrals under consideration, we will drop them from the notation in the following.

The second contribution is similar.
It also can be evaluated in terms of gamma functions at general $\eps$.
In the leading power, we obtain
\begin{align}
\clp{2,2}
=
\frac{\pi}{\eps}\left(\frac{v_{22}}{4}\right)^{-\eps}\frac{\Gamma(1+\eps)\Gamma(1-2\eps)}{v_{12}\,\Gamma(1-\eps)}\;.
\label{I2-2-res}  
\end{align} 

The third contribution is obtained in the leading power by setting $v_{11}=v_{22}=0$ under the integral sign:
\begin{align}
\regop{r_3} I_{1,1}\approxLP\frac{4 \sqrt{\pi}\,\Gamma(3/2-\eps)}{(1-2\eps)\Gamma(-\eps)} 
\int_0^\infty\dx t_1\int_0^\infty \dx t_2 \,
\frac{(1+t_2)^{-1/2 + \eps}\sqrt{t_2}}{(1 + t_1)^2 + 2 v_{12} t_1 t_2}\;.
\label{I2-3-int}
\end{align} 
To evaluate it one can apply the method of Mellin-Barnes (MB) representation\footnote{See, e.g., Chapter~5 of \cite{Smirnov:2012gma} or \cite{Dubovyk:2022} for a review.} by separating two terms in the denominator. Here a onefold representation is sufficient. The resulting MB integral is evaluated easily with an explicit result in an expansion up to the finite part in $\eps$
\begin{align}
\clp{2,3}=-\frac{2\pi}{v_{12}\,\eps}
+\frac{2\pi}{v_{12}}\log\frac{v_{12}}{2} +O(\eps)\;.
\label{I2-3-res}
\end{align} 
  
The sum of the three contributions to leading power and  to finite order in $\eps$ gives the following result
\begin{align}
I_{1,1} \approxLP \clp{2,1}+\clp{2,2}+\clp{2,3}= \frac{\pi}{v_{12}} \left(2\log v_{12}-\log \frac{v_{11}}{2} - \log \frac{v_{22}}{2} \right)+\mathcal{O}(\eps)\;.
\label{I2-res}
\end{align} 
The cancellation of poles in the sum  provides a useful standard check.
The result is in perfect agreement with the leading power approximation of the well-known full result for $I_{1,1}$ \cite{Schellekens:1981,Beenakker:1988,Somogyi:2011,Lyubovitskij:2021,Wunder:2024},
\begin{align}
I_{1,1}=\frac{\pi}{X}\log\left(\frac{v_{12}+\sqrt{X}}{v_{12}-\sqrt{X}}\right)+\mathcal{O}(\eps)\approxLP \frac{\pi}{v_{12}}\log\left(\frac{4 v_{12}^2}{v_{11}v_{22}}\right)+\mathcal{O}(\eps),
\end{align}
where $X=v_{12}^2-v_{11}v_{22}$.

Using expansion by regions of $I_{1,1}$, we obtained contributions from three regions, one proportional to $v_{11}^{-\eps}$, another proportional to $v_{22}^{-\eps}$, and one region independent of the masses.
A very different approach, based on an algebraic decomposition of the propagators, was employed in \cite{Wunder:2024}, also to extract the asymptotic behavior of $I_{1,1}$.
Interestingly, there the angular integral was also decomposed in precisely three parts.
This decomposition, using the two-point splitting lemma \cite{Lyubovitskij:2021,Wunder:2024}, is given by
\begin{align}
&I_{1,1}^{(2)}(v_{12},v_{11},v_{22};\eps)=\frac{1}{\sqrt{X}}\left[v_{34}\,I_{1,1}^{(0)}(v_{34};\eps)
-v_{13}\,I_{1,1}^{(1)}(v_{13},v_{11};\eps)
-v_{24}\,I_{1,1}^{(1)}(v_{24},v_{22};\eps)\right],
\label{eq: Splitting of double massive two denominator integral}
\end{align}
where the upper label in parentheses denotes the number of non-zero masses and kinematic invariants are
\begin{align}
X&=v_{12}^2-v_{11}v_{22}\approxLP v_{12}^2,\quad
v_{13}=\sqrt{X}\,\frac{\sqrt{X}+v_{11}-v_{12}}{2v_{12}-v_{11}-v_{22}}\approxLP\frac{v_{11}}{2}\,,\\
v_{23}&=\sqrt{X}\,\frac{\sqrt{X}+v_{22}-v_{12}}{2v_{12}-v_{11}-v_{22}}\approxLP\frac{v_{22}}{2}\,,\quad
v_{24}=\frac{2 X}{2v_{12}-v_{11}-v_{22}}\approxLP v_{12}
\end{align}
with ``$\approxLP$" indicating the leading power behavior for small masses $v_{11}, v_{22}$.
Therefore Eq.~\eqref{eq: Splitting of double massive two denominator integral}, approximated to leading power, simplifies to
\begin{align}
&I_{1,1}^{(2)}(v_{12},v_{11},v_{22};\eps)\approxLP\,I_{1,1}^{(0)}(v_{12};\eps)
-\frac{v_{11}}{2v_{12}}\,I_{1,1}^{(1)}\left(\frac{v_{11}}{2},v_{11};\eps\right)
-\frac{v_{22}}{2v_{12}}\,I_{1,1}^{(1)}\left(\frac{v_{22}}{2},v_{22};\eps\right).
\label{eq: Splitting of double massive two denominator integral LP}
\end{align}
We can immediately identify $I_{1,1}^{(0)}(v_{12};\eps)$ with the third region detected by \texttt{asy.m} since the contribution of Eq.~\eqref{I2-3-int} is exactly the angular integral $I_{1,1}$ with $v_{11}$ and $v_{22}$ set to zero, i.e. $\clp{2,3}=I_{1,1}^{(0)}$.

Using Eq.~(3.178) from \cite{Lyubovitskij:2021}, we have for the second contribution in Eq.~\eqref{eq: Splitting of double massive two denominator integral LP}
\begin{align}
I_{1,1}^{(1)}\left(\frac{v_{11}}{2},v_{11};\eps\right)=-\frac{2\pi}{\eps v_{11}}\left(\frac{v_{11}}{2}\right)^{-\eps}\mathrm{F}_1(-2\eps,-\eps,-\eps,1-2\eps;\omega_+,\omega_-)\,,
\end{align}
with $\mathrm{F}_1$ denoting the Appell function and
\begin{align}
\omega_\pm=1-\frac{1}{2}\,\frac{v_{11}}{1\pm\sqrt{1-v_{11}}}\,.
\end{align}
For small masses it is $\omega_+\approxLP 1$ and $\omega_-\approxLP 0$, hence the Appell function reduces to a product of Gamma functions resulting in
\begin{align}
-\frac{v_{11}}{2}I_{1,1}^{(1)}\left(\frac{v_{11}}{2},v_{11};\eps\right)\approxLP
\frac{\pi}{\eps}\left(\frac{v_{11}}{4}\right)^{-\eps}\frac{\Gamma(1+\eps)\Gamma(1-2\eps)}{v_{12}\,\Gamma(1-\eps)}=\clp{2,1}\,,
\end{align}
i.e. the second term of Eq.~\eqref{eq: Splitting of double massive two denominator integral LP} equals the contribution from the first region detected by \texttt{asy.m}.
Analogously, the contribution of the second region is found to be equal to the third term in Eq.~\eqref{eq: Splitting of double massive two denominator integral LP} at leading power.
Therefore expansion by regions and the algebraic decomposition based on the two-point splitting lemma lead to equivalent decompositions for the leading asymptotics.
Specifically, we found that each region of $I_{1,1}$ detected by \texttt{asy.m} can be interpreted as an angular integral at leading power.
Note that the scaling behavior of the angular integrals associated with the massive regions is different from the original angular integral since $v_{11}\sim v_{11}/2\approxLP v_{13}$, i.e. all kinematic arguments are small in the $I_{1,1}^{(1)}$ integrals in Eq.~\eqref{eq: Splitting of double massive two denominator integral} respectively \eqref{eq: Splitting of double massive two denominator integral LP}.

\subsection{Asymptotics for three denominators}

Let us now turn to the three-denominator angular master integral $I_{1,1,1}$ by evaluating the asymptotic behavior of the integral representation (\ref{eq: Three-denominator angular integral asym}) with all three indices equal to one
\begin{align}
I_{1,1,1}=-(2 \eps+1)\pi \int_0^\infty\!\!\dx t_1\int_0^\infty \!\!\dx t_2\int_0^\infty \!\!\dx t_3\,.
  (1+t_2)t_3 (1+t_3)^{\eps} (w_3(t_1,t_2,t_3))^{-3/2}\;.
\label{I3}
\end{align}
The massless limit is $v_{11}\sim v_{22}\sim v_{33} \ll v_{ij}$ for $i\neq j$.
The polynomial factors $(1+t_2) t_3$ are irrelevant as far as the set of regions contributing to the expansion
is concerned. To reveal regions we apply the following command:
\begin{verbatim}
 r = WilsonExpand[(1+t[3]),(1+t[1])^2 (1+t[2])^2 
    + (v33 y + 2 v23 t[2] + v22 y t[2]^2 + v11 y t[1]^2 (1 + t[2])^2 
    + 2 t[1](1 + t[2]) (v13 + v12 t[2])) t[3], {t[1],t[2],t[3]}, {y -> x}]
\end{verbatim}
Here $y$ is again the small parameter of the problem. As an output of this command, we obtain 
\begin{align}
r =\{\{1, -1,  -1\}, \{1, 1, -1\}, \{-1, 0, -1\}, \{0, 0, 0\}\}.
\end{align}

The contribution of a given region $r_i$ is again obtained by
the scaling $t_j \to y^{(r_i)_j} t_j$, multiplying by  $y^{\sum_{j=1}^3 (r_j)_i}$, expanding in $y$ at $y=0$ and setting $y=1$ in the end.
We obtain the following leading order behavior of the first contribution:
\begin{align}
\regop{r_1} I_{1,1,1}\approxLP-(1+2 \eps)\pi \int_0^\infty\!\!\dx t_1\int_0^\infty \!\!\dx t_2\int_0^\infty \!\!\dx t_3\,
 \frac{ t_2 t_3^{1+\eps}}{ (2 t_1 t_2^2 t_3 v_{12} + t_2 (t_2 + t_2 t_3 v_{22} + 2 t_3 v_{23}) )^{3/2}}\;.
\label{I3-1}
\end{align}
The integral can be easily evaluated by an application of a one-dimensional integration formula, with the following result
valid at general $\eps$:
\begin{align}
\clp{3,1}=
\frac{\pi}{\eps}\left(\frac{v_{22}}{4}\right)^{-\eps}\frac{\Gamma(1+\eps)\Gamma(1-2\eps)}{v_{12}v_{23}\,\Gamma(1-\eps)}\,.
\label{I3-1-res}
\end{align}
The second and the third contributions are similar. In the leading power, we obtain
\begin{align}
\clp{3,2}=
\frac{\pi}{\eps}\left(\frac{v_{33}}{4}\right)^{-\eps}\frac{\Gamma(1+\eps)\Gamma(1-2\eps)}{v_{13}v_{23}\,\Gamma(1-\eps)}\,,
\label{I3-2-res}\\
\clp{3,3}=
\frac{\pi}{\eps}\left(\frac{v_{11}}{4}\right)^{-\eps}\frac{\Gamma(1+\eps)\Gamma(1-2\eps)}{v_{12}v_{13}\,\Gamma(1-\eps)}\,.
\label{I3-3-res}
\end{align}

The fourth contribution is obtained by expanding the integrand of (\ref{I3}) in a Taylor series in $v_{ii}$, $i=1,2,3$.
In the leading order, the corresponding contribution is given by Eq.~\eqref{I3} with $w_3$ replaced by
\begin{align}
w_{30}(t_1,t_2,t_3)&=(1+t_1)^2(1+t_2)^2 +\left[ 2 t_1 (1+t_2)(v_{13}+t_2 v_{12}) +2 t_2 v_{23} \right]t_3\,.
\end{align}
We evaluate it by the method of Mellin-Barnes (MB) representation. To derive an appropriate MB representation one can apply the public computer program {\tt MBcreate} \cite{Belitsky:2022gba}.
We used the following MB representation:
\begin{align}
\regop{r_4} I_{1,1,1}\approxLP - 2\sqrt{\pi}\,\frac{1 + 2 \eps}{\Gamma(-\eps)}
  \frac{1}{(2\pi i)^3} \int_{-i\infty}^{i\infty}\!\!\dx z_1\int_{-i\infty}^{i\infty} \!\!\dx z_2\int_{-i\infty}^{i\infty} \!\!\dx z_3\, 
2^{z_1} v_{12}^{z_3} v_{13}^{z_2 - z_3} v_{23}^{z_1 - z_2} \nonumber \\
\times \frac{\Gamma(3/2 + z_1)\Gamma(2 + z_1)}{\Gamma(3 + 2 z_1)} \Gamma(-2 - \eps - z_1)
\Gamma(1 + z_2) 
\nonumber \\
\times
\Gamma(z_2-z_1) 
\Gamma(1 + z_1 - z_3) \Gamma(-z_3) \Gamma(z_3-z_2) \Gamma(1 + z_1 - z_2 + z_3)\;,
\label{I3-4-MB}
\end{align}
where the integration contours are chosen in such a way that the poles of gamma functions with $-z$-dependence are to the
right of the contour and the poles   with $+z$-dependence are to the
left of the contour \cite{Smirnov:2012gma}. To convert a given MB  integral into a form where
an expansion in $\eps$ is possible under the integral sign, one can apply the two public codes {\tt MB.m} and {\tt MBresolve.m}
\cite{Czakon:2005rk,Smirnov:2009up}. After such a resolution of singularities is performed,
one can apply the command {\tt DoAllBarnes} from Kosower's\footnote{All required MB tools 
can be downloaded from {\tt https://gitlab.com/feynmanintegrals/mb}.} {\tt barnesroutines.m} which
automatically applies the first and the second Barnes lemmas (and their corollaries).
 
We arrive at the following result for the fourth contribution in an $\eps$-expansion up to the finite part: 
\begin{align}
\clp{3,4}&=-\frac{\pi  (v_{12}+v_{13}+v_{23})}{\eps\,v_{12} v_{13} v_{23}}
-\frac{\pi}{v_{12}v_{13}v_{23}}\left[
(v_{12}-v_{13}-v_{23})\log\frac{v_{12}}{2}
\right.
\nonumber\\
&\qquad\qquad\qquad
\left.
+(v_{13}-v_{12}-v_{23})\log\frac{v_{13}}{2}
+(v_{23}-v_{12}-v_{13})\log\frac{v_{23}}{2}
\right]+\mathcal{O}(\eps)\,.
    \label{I3-4-res}
\end{align}
We note that this $\eps$-expansion agrees with the $\eps$-expansion of the massless three-denominator integral
\footnote{Mind the different normalization factors for the $v_{ij}$ in the Somogyi paper, use $v_{ii}\rightarrow 4 v_{ii}$ respectively $v_{ij}\rightarrow 2 v_{ij}$ for $i\neq j$ to get from Eq.~\eqref{I3-4-res} to Eq.~(C46) of \citep{Somogyi:2011}.} \citep{Somogyi:2011}.
It turns out that the poles in the sum of the first three contributions (\ref{I3-1-res}), (\ref{I3-2-res}), (\ref{I3-3-res}),
are cancelled with the pole part (\ref{I3-4-res}) of the fourth contribution which is a natural check of the expansion procedure.

We obtain the following result for the leading power in an $\eps$-expansion up to the finite part: 
\begin{align}
I_{1,1,1} &\approxLP \clp{3,1}+\clp{3,2}+\clp{3,3}+\clp{3,4}\\
\nonumber
&=\frac{\pi}{v_{12} v_{13} v_{23}}
\left[
(-v_{12}+v_{13}+v_{23})\log v_{12}
+(v_{12}-v_{13}+v_{23})\log v_{13}
\right.
\nonumber
\vphantom{\frac{X}{X}}
\\
&\qquad\left.
+(v_{12}+v_{13}-v_{23})\log v_{23}
- v_{23} \log\frac{v_{11}}{2}
- v_{13} \log\frac{v_{22}}{2}
- v_{12} \log\frac{v_{33}}{2}
\right]
+\mathcal{O}(\eps).
   \label{eq: I111 LP res}
\end{align}
Let us emphasize that the sum of the first three contributions provides all the leading logarithms of the small parameters
$v_{ii}$ in any order of $\eps$.

Eq.~\eqref{eq: I111 LP res} constitutes a new result for the fully massive three-denominator angular integral.
Results with some masses set to zero are easily obtained by switching off the corresponding region scaling like $v_{ii}^{-\eps}$.
Hence we have
\begin{align}
&I_{1,1,1}^{(2)}(v_{12},v_{13},v_{23},v_{11},v_{22})\approxLP \clp{3,1}+\clp{3,3}+\clp{3,4}\nonumber\\
&=-\frac{\pi}{\eps\,v_{13}v_{23}}+\frac{\pi}{v_{12} v_{13} v_{23}}\left[
(-v_{12}+v_{13}+v_{23})\log v_{12}
+(v_{12}-v_{13}+v_{23})\log v_{13}
\right.
\nonumber
\vphantom{\frac{X}{X}}
\\
&\qquad\left.
+(v_{12}+v_{13}-v_{23})\log v_{23}
- v_{23} \log\frac{v_{11}}{2}
- v_{13} \log\frac{v_{22}}{2}
- v_{12} \log 2
\right]
+\mathcal{O}(\eps)
\end{align}
and 
\begin{align}
&I_{1,1,1}^{(1)}(v_{12},v_{13},v_{23},v_{11})\approxLP \clp{3,3}+\clp{3,4}\nonumber\\
&=-\frac{\pi(v_{12}+v_{13})}{\eps\,v_{12}v_{13}v_{23}}+\frac{\pi}{v_{12}v_{13}v_{23}}\left[
(-v_{12}+v_{13}+v_{23})\log v_{12}
+(v_{12}-v_{13}+v_{23})\log v_{13}
\right.
\nonumber
\vphantom{\frac{X}{X}}
\\
&\qquad\left.
+(v_{12}+v_{13}-v_{23})\log v_{23}
- v_{23} \log\frac{v_{11}}{2}
- v_{13} \log 2
- v_{12} \log 2
\right]
+\mathcal{O}(\eps).
\end{align}
As noted above, the massless integral is identical to the contribution if the region $\clp{3,4}$, i.e. $I_{1,1,1}^{(0)}(v_{12},v_{13},v_{23})=\clp{34}$ with its $\eps$-expansion given in Eq.~\eqref{I3-4-res}.
In all three cases there remain collinear poles originating from the massless region.
The presented results for the massive three denominator integrals, especially the leading logarithms in $v_{ii}$ to all-orders in $\eps$, provide a valuable check for potential future calculations of the yet unknown higher orders in $\eps$ of the three-denominator angular integral $I_{1,1,1}$.

\subsection{Asymptotics for four denominators}
\label{sec: Four denominators asy}
 
We continue our discussion with the four denominator angular master integral $I_{1,1,1,1}$.
For this, let us evaluate the asymptotic behavior of the integral representation (\ref{eq: Four-denominator angular integral asym}) with all the four indices equal to one in the massless limit
\begin{align}
I_{1,1,1,1}=&
\frac{16\sqrt{\pi}}{1-2\eps}\frac{\Gamma \left(\frac{3}{2}-\eps\right)}{ \Gamma\left(-\eps-1\right)}\nonumber\\
&\times
\int_0^\infty\!\dx t_1\int_0^\infty\!\dx t_2\int_0^\infty\!\dx t_3\int_0^\infty\!\dx t_4\,(1+t_2)
   (1+t_3)^2 t_4^{\frac{3}{2}}
   (1+t_4)^{\eps+\frac{1}{2}} w_4^{-2}\;.
\label{I4}
\end{align}
 
To reveal regions contributing in the massless limit we apply  \texttt{asy.m} with
\begin{verbatim}
r= WilsonExpand[(1 + t[4]) t[4], (1 + t[1])^2 (1 + t[2])^2 (1 + t[3])^2 
+ (v44 y + 2 v34 t[3] + v33 y t[3]^2 + v22 y t[2]^2 (1 + t[3])^2 
+ v11 y t[1]^2 (1 + t[2])^2 (1 + t[3])^2 
+ 2 t[2] (1 + t[3]) (v24 + v23 t[3]) 
+ 2 t[1] (1 + t[2]) (1 + t[3]) (v14 + v13 t[3] 
+ v12 t[2] (1 + t[3]))) t[4], {t[1], t[2], t[3], t[4]}, {y -> t}]
\end{verbatim}
 As an output we obtain 
\begin{align}
r =\{\{1, 1, -1, -1\}, \{1, 1, 1, -1\}, \{0, 0, 0, 0\}, \{1, -1, 0, -1\}, \{-1, 0, 0, -1\}\}\;.
\end{align} 
 
For the leading power of the first contribution we find
\begin{align}
&\regop{r_1}I_{1,1,1,1}\approxLP 4 \sqrt{\pi}\,\frac{\Gamma(1/2 - \eps)}{3\,\Gamma(-1 - \eps)}  
  \nonumber\\
&\times \int_0^\infty\!\!\dx t_1\int_0^\infty \!\!\dx t_2\int_0^\infty \!\!\dx t_3\int_0^\infty \!\!\dx t_4\,
          \frac{t_4^{2 + \eps}}{(t_3 + 2 t_1 t_3 t_4 v_{13} + 2 t_2 t_3 t_4 v_{23} + t_3 t_4 v_{33} + 2 t_4 v_{34})^2}\;.
\label{I4-1}
\end{align}
As in the case of three denominators, such integral can be evaluated by an application of a one-dimensional 
integration formula, with the following result:
\begin{align}
\clp{4,1}=
\frac{\pi}{\eps}\left(\frac{v_{33}}{4}\right)^{-\eps}\frac{\Gamma(1+\eps)\Gamma(1-2\eps)}{v_{13}v_{23}v_{34}\,\Gamma(1-\eps)}\,.
\label{I4-1-res}
\end{align}
  
The contributions of regions with numbers 2,4 and 5 can similarly be evaluated with the following results:
\begin{align}
\clp{4,2}=
\frac{\pi}{\eps}\left(\frac{v_{44}}{4}\right)^{-\eps}\frac{\Gamma(1+\eps)\Gamma(1-2\eps)}{v_{14}v_{24}v_{34}\,\Gamma(1-\eps)}\,,
\label{I4-2-res}
\\
\clp{4,4}=
\frac{\pi}{\eps}\left(\frac{v_{22}}{4}\right)^{-\eps}\frac{\Gamma(1+\eps)\Gamma(1-2\eps)}{v_{12}v_{23}v_{24}\,\Gamma(1-\eps)}\,,
\label{I4-4-res}
\\
\clp{4,5}=
\frac{\pi}{\eps}\left(\frac{v_{11}}{4}\right)^{-\eps}\frac{\Gamma(1+\eps)\Gamma(1-2\eps)}{v_{12}v_{13}v_{14}\,\Gamma(1-\eps)}\,.
\label{I4-5-res}
\end{align}
  
The third contribution is obtained by expanding the integrand of (\ref{I4}) in a Taylor series in $v_{ii}$, $i=1,2,3,4$.
In the leading power, the corresponding contribution is given by (\ref{I4}) with $w_4$ replaced by
the corresponding expression at $v_{ii}=0$.
To evaluate it one can apply the method of Mellin-Barnes (MB) representation like in the previous case.
Here the program {\tt MBcreate} provides a sixfold MB representation. 
The result for the pole part, which we obtained with the same MB tools characterized in the previous subsection, is
\begin{align}
%
\clp{4,3}=-\frac{\pi}{\eps}\left(\frac{1}{v_{12}v_{13}v_{14}}+\frac{1}{v_{12}v_{23}v_{24}}+\frac{1}{v_{13}v_{23}v_{34}}+\frac{1}{v_{14}v_{24}v_{34}}\right)+\mathcal{O}\!\left(\eps^0\right).
\label{I4-3-res}
\end{align}
It turns out that it cancels the sum of the pole parts of the previous
four contributions providing an important check.

However, the finite part in $\eps$ of the third contribution is more complicated. 
It involves MB integrals with up to four integrations.
Presumably, it is reasonable to look for a more appropriate method for its analytic evaluation.
Still like in the previous case, the sum of the other contributions \eqref{I4-1-res}-\eqref{I4-5-res} contains all the leading logarithmical asymptotics in any order of the $\eps$-expansion.
This statement holds not only for the leading power contribution but also for any higher power contribution in the asymptotic expansion in the masses.
The corresponding integrals can be taken explicitly, exactly in the same way that was used for the leading power contribution.

As a consistency check for the result obtained for the pole of the massless four-denominator integral $I_{1,1,1,1}^{(0)}$, we can look at the special case where the spatial parts of the vectors $(\mathbf{v}_i)_{i=1,\dots,4}$ are linearly dependent, i.e. there are non-zero $\lambda_i$ such that $0=\sum_i \lambda_i \mathbf{v}_i$.
Note that this assumption is necessarily satisfied if the vectors $v_i$ are confined to a $4$ dimensional subspace of the $d$ dimensional space, which is the case in many physical applications.
Under the assumption of linear dependence we have the reduction identity \cite{Lyubovitskij:2021}
\begin{align}
I_{1,1,1,1}(v_1,v_2,v_3,v_4)=\frac{1}{\sum_i \lambda_i}&\left[
\lambda_1 I_{1,1,1}(v_2,v_3,v_4)+
\lambda_2 I_{1,1,1}(v_1,v_3,v_4)
\right.\nonumber\\
&\left.+
\lambda_3 I_{1,1,1}(v_1,v_2,v_4)+
\lambda_4 I_{1,1,1}(v_1,v_2,v_3)
\right].
\label{eq: I1111 lin dep decomposition}
\end{align}
Considering all vectors massless, the pole parts of the three-denominator integrals are, as we can read of from Eq.~\eqref{I3-4-res},
\begin{align}
I_{1,1,1}^{(0)}(v_{12},v_{13},v_{23})=-\frac{\pi}{\eps}\left(\frac{1}{v_{12} v_{13}}+\frac{1}{v_{12} v_{23}}+\frac{1}{v_{13} v_{23}}\right)+\mathcal{O}\!\left(\eps^0\right)\,.
\end{align}
Plugging this into Eq.~\eqref{eq: I1111 lin dep decomposition} and sorting by denominators we observe that the $\lambda_i$ dependence cancels out and we recover exactly the result from Eq.~\eqref{I4-3-res}.
This provides a consistency check for the contribution of the massless region $\clp{4,3}$. Furthermore, we observe that the structure of the pole of $I_{1,1,1,1}^{(0)}$ does not depend on whether the involved vectors are linearly dependent.

\subsection[General structure of the leading asymptotics for $n$ denominators]{General structure of the leading asymptotics for \boldmath{$n$} denominators}
Looking at the regions found for two, three, and four denominators, we observe a rather simple pattern.
The following statements have been shown explicitly in the cases $n\leq 4$ in the preceding sections and conjecturally hold true for an arbitrary number of denominators.
A formal proof for general $n$ is beyond the scope of this paper.

For the $n$-denominator angular integral $I_{1,\dots,1}(v_1,\dots,v_n)$ there are $n+1$ regions contributing to its leading asymptotics.
Up to relabeling of the regions, there is one region $r_i$ associated with each massive vector $v_i$ which contributes with
\begin{align}
\clp{n,i}
=\frac{\pi}{\eps}\left(\frac{v_{ii}}{4}\right)^{-\eps}\frac{\Gamma(1+\eps)\Gamma(1-2\eps)}{\Gamma(1-\eps)}\prod_{\overset{j=1}{j\neq i}}^n \frac{1}{v_{ij}}
\end{align}
and one mass-independent region $r_{n+1}$ which is equal to the massless $n$-denominator integral
\begin{align}
\clp{n,n+1}=I_{1,\dots,1}^{(0)}\,.
\end{align}
The leading power approximation of the $n$-denominator angular integral is therefore given by
\begin{align}
I_{1,\dots,1}(v_1,\dots,v_n)&\approxLP \sum_{i=1}^{n+1} \clp{n,i}=I_{1,\dots,1}^{(0)}+\frac{\pi}{\eps}\frac{\Gamma(1+\eps)\Gamma(1-2\eps)}{\Gamma(1-\eps)}\sum_{i=1}^n \left(\frac{v_{ii}}{4}\right)^{-\eps}\prod_{\overset{j=1}{j\neq i}}^n \frac{1}{v_{ij}}\,.
\label{eq: conjecture for n denominators}
\end{align}
If all $n$ masses $v_{ii}$ are non-zero we know that $I_{1,\dots,1}$ is free of collinear singularities, hence the $\eps$-pole will cancel between the massive regions and the massless regions.
Therefore Eq.~\eqref{eq: conjecture for n denominators} implies for the pole of the $n$-denominator massless integral
\begin{align}
I_{1,\dots,1}^{(0)}(v_1,\dots,v_n)=-\frac{\pi}{\eps}\sum_{i=1}^n\prod_{\overset{j=1}{j\neq i}}^n \frac{1}{v_{ij}}+\mathcal{O}\!\left(\eps^0\right).
\label{eq: n denominators massless pole}
\end{align}
For the case of $m$ masses, say $v_{11},\dots,v_{mm}\neq 0$ and $v_{m+1,m+1},\dots,v_{nn}=0$, the contributions of the regions $r_{m+1},\dots,r_n$ vanish.
Hence we get for the pole of the $n$-denominator integral with $m$ masses
\begin{align}
I_{1,\dots,1}^{(m)}(v_1,\dots,v_n)&\approxLP \clp{n,n+1}+\sum_{i=1}^{m} \clp{n,i}
=-\frac{\pi}{\eps}\sum_{i=m+1}^{n}\prod_{\overset{j=1}{j\neq i}}^n \frac{1}{v_{ij}}+\mathcal{O}\!\left(\eps^0\right).
\end{align}
These predictions for the pole structure for general $n$ as well as the leading asymptotics for small masses may serve as useful checks for future analytic calculations of multi-denominator angular integrals.
\section{Conclusion}
\label{sec: Conclusion}
We have successfully applied the method of expansions by regions to angular integrals.
Using the code \texttt{asy.m}, we calculated the leading asymptotic behavior of angular integrals in the massless limit.
For the calculation of the contributions of massless regions Mellin-Barnes representations have been applied.
We established explicit results for the previously unknown cases of three and four denominators.
In both cases, we find the expected cancellation of poles and obtain the leading logarithmic contributions in all orders of the $\eps$-expansion.
Based on these results we formulated a conjecture for the leading asymptotics and pole structure for a general number of denominators and masses.
These findings may serve as useful checks for future calculations of multi-denominator angular integrals.

Furthermore we observe an interesting connection between the regions detected by \texttt{asy.m} and an algebraic decomposition of angular integrals.
This allows for an identification of regions for the angular integrals in terms of angular integrals with a different scaling behavior.

\acknowledgments

The work of V.S. was supported by the Russian Science Foundation under the agreement
No. 21-71-30003 (applications of the method of MB representation) and by the Ministry
of Education and Science of the Russian Federation as part of the program of the Moscow
Center for Fundamental and Applied Mathematics under Agreement No. 075-15-2022-284
(applications of expansion by regions).

\bibliography{AiEr}
\bibliographystyle{JHEP}
\end{document}